# Designing Spontaneous Speech Search Interface for Historical Archives

Donna Vakharia
Department of Computer Science
University of Texas at Austin
donna@cs.utexas.edu

Rachel Gibbs
School of Information
University of Texas at Austin
rachel.gibbs@utexas.edu

## ABSTRACT
Spontaneous speech in the form of conversations, meetings, voice-mail, interviews, oral history, etc. is one of the most ubiquitous forms of human communication. Search engines providing access to such speech collections have the potential to better inform intelligence and make relevant data over vast audio/video archives available to users. This project presents a search user interface design supporting search tasks over a speech collection consisting of an historical archive with nearly 52,000 audiovisual testimonies of survivors and witnesses of the Holocaust and other genocides. The design incorporates faceted search, along with other UI elements like highlighted search items, tags, snippets, etc., to promote discovery and exploratory search. Two different designs have been created to support both manual and automated transcripts. Evaluation was performed using human subjects to measure accuracy in retrieving results, understanding user-perspective on the design elements, and ease of parsing information.

## 1. INTRODUCTION
The ease of capturing and encoding of audiovisual data has resulted in the production of a massive speech collection with limited tools to analyze the data effectively. As the magnitude and use of such content grows, efficient ways to automatically find relevant segments becomes necessary. While there are efficient search engines for text documents present today, there are no satisfactory systems for performing search over audiovisual data as these systems do not perform any detailed analysis for them [11].

We know that core audio-video dimension of oral history is notoriously underutilized [3]. Historical archives provide rich information through the vast set of data. Even for the properly catalogued audiovisual collection, all the efforts in collecting this dataset would be of little value if they fail to be easily accessible by the target audience. These collections would be rendered unlistened and unwatched in such a scenario. The content is rarely organized, indexed or searchable/browsable in any useful way. Because of which the considerable potential of audio and video documents to support high-impact, vivid, thematic, and analytic engagement with meaningful issues, personalities, and contexts, is largely untapped [3]. These archives can be made usable by representing them in text by either manual transcription which is quite expensive. Or automated methods or Automatic speech recognition (ASR) tools can be used to convert to text. Text is easier to read, scan, browse, search, publish, display, and distribute. Audio or video documents, in contrast, inevitably have to be experienced in real time.

The Survivors of the Shoah Visual History Foundation (VHF) has created an archive that contains interviews of nearly 52,000 survivors and other witnesses of the Holocaust. The institute interviewed Jewish, homosexual, and Jehovah's Witness, and Roma and Sinti (Gypsy) survivors in addition to liberators and liberation witnesses, political prisoners, rescuers and aid providers, survivors of Eugenics policies, and war crimes trials participants [8]. Approximately 25000 of the collected interviews are in English, 7000 are in Russian, and 575 are in Czech. Presently USC's Visual History Archive Online enables researchers to perform detailed searches for relevant testimonies and segments of testimonies but with no access to the transcripts.

This study presents a search system using the VHF collection that enables users to perform content-based search and retrieve the transcript of the audio relevant to the query. Since manual transcription can be very expensive and time consuming, we try to evaluate the alternative solution by using auto-generated transcripts. However, since transcripts generated by ASR are usually of bad quality and disregards disfluencies, misspellings, sentence end annotations, speaker change annotations, etc. we harness the use of various design elements to enhance readability of the transcripts. We further evaluate user efficiency in finding relevant results using them as opposed to manual transcripts.

### 1.1 Proposed Solution
- Design a search system that would enable users to perform content-based search and to retrieve the transcript of the interview relevant to the query.
- Support search tasks over the content vs. the tags, keywords associated with the file.
- Support faceted search to help users perform exploratory search and enable discovery.

- Utilize elements such as keyword highlighting, snippets, category search, tags, etc., to enhance user experience.

### 1.2 Intended users

The intended users for the system are majorly historians, academicians, researchers, genealogist, students, and history-enthusiasts. This intended user group includes people of various levels of domain and technical expertise, from experts to novices. Since the collection focuses on a topic of global interest, users will vary in terms of native language, knowledge of terminology, and education levels.

### 1.3 Supported search tasks

Exploratory and Iterative Search, Re-finding, and Known Item Search will be supported, but Known Item Search will be less emphasized as it will be a less-common use case.

The remainder of the paper is organized as follows. Section 2 explains the motivation behind this study. Prior work on data evaluation of design of search systems is briefly reviewed in Section 3. Interface design elements are described in Section 4. Section 5 enlists our research hypothesis that we test using pre-defined methodology, presented in Section 6. Evaluation is performed on human subjects and results obtained from them in terms of self-reported feedback, and calculated measures is presented in Section 7. Finally, Section 8 discusses the research areas that need further investigation in future, followed by conclusion in Section 9.

## 2. MOTIVATION

The researchers were motivated to design for this data set for a number of reasons. Both researchers had previously studied the subject extensively and had personal connections to the subject through either cultural exposure or familial relation to survivors and victims. One researcher was already involved in a research activity using this same data set, and a relative's interview of the other was included in this survivor-testimony collection, conducted by the USC Shoah Foundation Institute. Therefore, both researchers had an interested in increasing access to the audiovisual interviews and transcripts. This collection is extensive and was collected as quickly as possible to preserve the first-hand accounts of witnesses and survivors before they were lost forever.

The ease of capture and encoding of audiovisual data has caused a massive amount of speech collection to be produced with limited tools to analyze them effectively. While there are efficient search engines for text documents present today, there are no satisfactory systems for performing search over audiovisual data. This project sought to design a search system that would enable users to perform content-based search and to retrieve the transcript of the interview relevant to the query. Due to the vastness of the collection, human transcription is laborious, time-consuming, and expensive. For this reason, alternate methods of making the texts of the collection available are explored, such as automatic transcription using a variety of algorithms. In addition to the elements of the user interface improving accessibility to the collection, the readability of transcriptions must also be considered and tested. The design of this interface and testing of user comprehension of automatic transcription against human transcription seeks to accommodate any user with access to the collection, from expert researchers to survivors or their families to students seeking to learn and discover about the tragic event upon which this collection focuses.

## 3. RELATED WORK

Various studies and experiments have been performed to address the problem of finding relevant information from vast oral/visual archives. Jones et al. [2] performed a psycholinguistic study measuring readability of several types of speech transcripts. They gauged readability by measuring accuracy of answers to comprehension questions, reaction-time for passage reading, reaction-time for question answering and a subjective rating of passage difficulty. However, the data source for their study included texts from conversational telephone speech, and broadcast news speech and was based completely on automatic transcripts. A study comparing the effects of manual vs. automatic transcripts still remains unexplored.

The Informedia Digital Video Library Project at Carnegie Mellon University is creating a digital library of text, images, videos and audio data available for full content retrieval [4]. Hauptmann et al. built a system by integrating technologies involved in creating a digital video library suitable for full-content search and retrieval [6]. They use image processing to analyze scenes, speech processing to transcribe audio signal, and natural language processing to determine word relevance. Doing so, they try to overcome limitations of each technology [5].

Coden et al. also try to address the problem of finding pertinent information for visual data [7]. However their problem includes performing this in real time. To tackle the issue with having transcripts with low accuracy they develop algorithms to extract the essence of video and use distance-ranking to select relevant information from the search results. Their evaluation involved data analysis using quantitative methods alone. Also, their system is designed to handle broadcast news data. Such automated techniques to search broadcast news are difficult to apply for larger collections of spontaneous speech.

An experimental speech-based search engine, SpeechBot, uses automatic transcripts for indexing and in turn uses searchable index to provide an ability to play the segments of interest within the audio file [11]. Many other studies [4, 13, and 14] have been carried out to work towards similar goals.

## 4. INTERFACE DESIGN

When ideating for the search user interface (SUI) design, the researchers considered likely user types and the types of information needs and levels of domain expertise (or lack-there-of) of each group. Since the purpose of the interface is to increase access to the transcripts of audiovisual interviews with survivors and witnesses of the Holocaust and other genocides, a widely studied topic, the user types varied greatly and included: historians/scholars, teachers (any level), students (any level), genealogists, and anyone who has an interest in studying primary source material for this topic. To accommodate users with different levels of domain expertise, the SUI utilized various elements to support information retrieval and discovery. The desired elements were identified and explored during an iterative design process.

In addition to a summary about the searchable collection, the SUI landing page presents several ways in which users can instigate search. For the search bar, an autosuggest feature (figure 1) was especially useful as the data set includes many non-English terms and terms with numerous spellings.

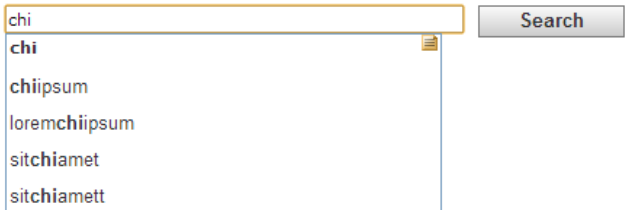

**Figure 1: Auto-suggest**

While the search bar serves to instigate search for users with a known information need, also introduced on the landing page are a several search elements which persist throughout the interface to support exploratory search where the information need is unknown or unclear. These include: Popular Searches (determined by what other users search), Suggested Searches (curated search list), Featured Interview that contained dynamic excerpt from transcription text (figure 2), and a Browse By category filter (also used as faceted search).

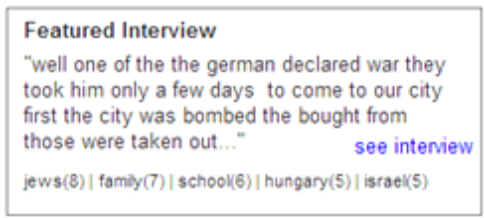

**Figure 2: Featured interview box is present on all the pages of the interface to instigate user interest**

Arriving on a search engine results page (SERP), users could take several options:

(1) select/explore the search results presented on the page, (2) narrow their search using Time Period (1930-1938, 1939-1945, 1946-present), Location (Western Europe, Central Europe, Eastern Europe, Southern Europe, Scandinavia), or Subject filters (Kristallnacht, Resettlement & deportation, Concentration & labor camps, Ghettos, Pogroms, Persecuted groups, Death squads, Extermination camps, Jewish resistance, Escapes, Death marches, Liberation) from the left hand facets, (3) reformulate the existing query which persisted in the search bar, or (4) start a new search by selecting the featured interview, a Popular Searches topic, or a topic from Suggested Searches.

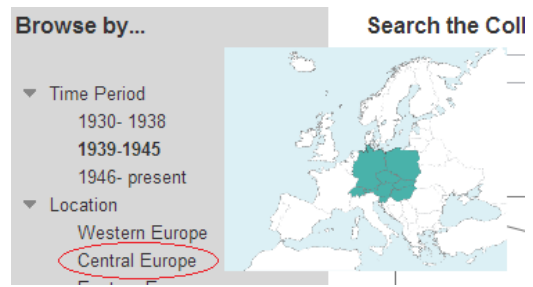

**Figure 3: Maps appear on mouse over on location filters to assist users**

The search results listings presented several different elements to assist users in evaluating the perceived relevancy of results to their information needs. These included: a hyperlinked title (with descriptiveness dependent upon whether the associated result was human or machine-transcribed), descriptive text (either human-created summary or excerpt from the interview transcript) with highlighted relevant words (either from search query terms or category filters), a snippet discoverable upon mouseover (figure 4) that included a longer excerpt from the transcript with more keyword highlights, search result listings also offered users a hyperlink to go directly to audio or video recording of the interview. All of these features sought to help users evaluate whether or not a particular result was likely to match the information need as much as possible before clicking on and evaluating the result itself.

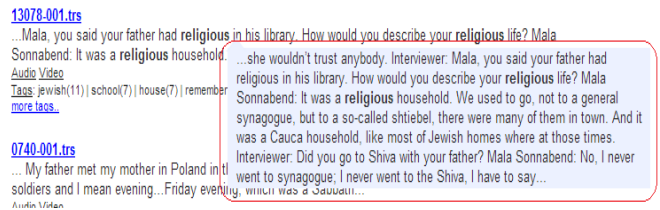

**Figure 4: Snippets with keyword highlighting**

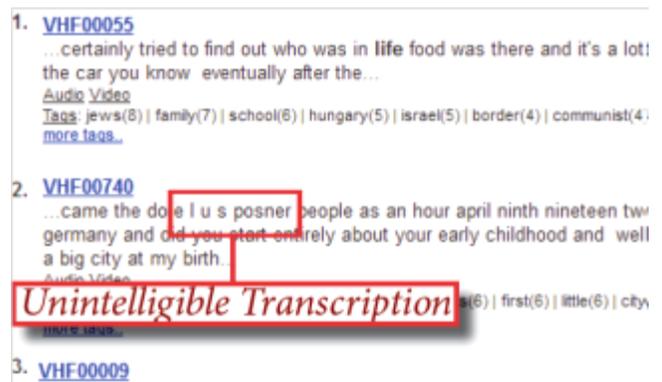

**Figure 5: Sample SERP containing unintelligible transcription**

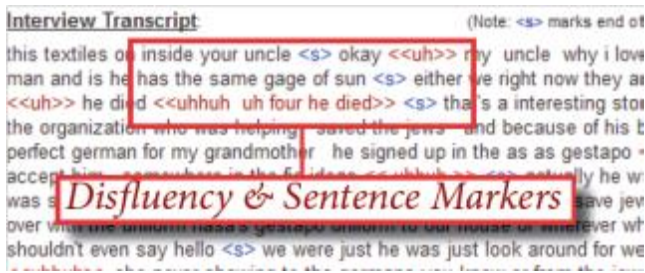

**Figure 6: Sample result using ASR text with disfluency and end-of-sentence markers**

Once clicked, the result presented either a human or automatic transcription of a witness or survivor testimony [please see Appendix for complete screenshots]. Certain elements were presented on both versions of transcripts such as back navigation to the search results, a list of frequency tags, and links to the audio and video recordings which could be utilized as a companion to the transcriptions. The formatting and readability of the transcription text themselves varied depending on the method of transcription. Human-transcribed texts were more readable, presented clear indications of speaker changes. Machine transcripts presented color-coded spontaneous speech markup such as disfluency, sentence, and speaker markers (figure 5, 6).

## 5. RESEARCH HYPOTHESES

The study sought to evaluate the proposed designs of the SUI via a user study. This study was conducted at the University of Texas at Austin, School of Information with a convenience sample of 8 users of mixed ethnographic backgrounds, levels of domain expertise, and familiarity with database/archival search interfaces. The study sought to test 3 hypotheses regarding the users' interactions and perceptions of the presentation of information in the SUI and search result documents:

Hypothesis 1 (H1): Users will prefer human-transcribed search results, as opposed to those generated automatically by machine.

Hypothesis 2 (H2): As the level of disfluency is reduced and sentence markers are added, users will be able to judge perceived relevance of presented search results more quickly and with greater confidence.

Hypothesis 3 (H3): Marking speaker changes within the text of the transcripts will improve user understanding and search efficiency, leading users to find desired results in a shorter amount of time (contrasted with transcriptions without marked speaker changes).

## 6. METHODOLOGY

In this study, users were guided through a short series of tasks designed to gain feedback regarding search user interfaces and clarity of information presentation. Tasks focused on describing impressions, understanding of information architecture, entering pre-defined search queries, refining search queries, and evaluating the ease of parsing the presented search artifacts. Morae[1] software was used to capture user interactions during the evaluation phase via screen, and audio recording for Tasks 1, 2, and 3. The evaluation phase was divided into following tasks (figure 7).

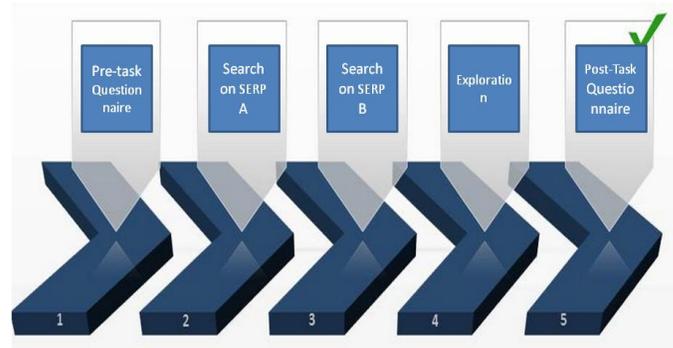

**Figure 7: Evaluation Steps**

**Task 0 – Signing pre-consent form and completing pre-task questionnaire:** All users signed an informed consent form and were given a copy for their records. Users completed a digital questionnaire collecting information concerning ethnographic details and educational background before beginning the actual tasks. No personally identifiable information was recorded during the study.

*Objectives*: To gather information regarding user's ages, educational backgrounds, ethnographic details, and gauge level of familiarity with search engines, archives, and prior experience researching on Holocaust (if any).

**Task 1 – Search by keyword using SERP A**: Users were assigned a search task to find a relevant result for the topic "*Pre-war religious festival observance*". Although not informed of the data source for the transcriptions of this search task, Task 1 presented users with the SERP from "Good" or human-transcribed search results. Users were then given the following scenario for their information need:

*You are trying to find information about how religious festivals in pre-war Europe were observed. You are particularly interested in observance of three main festivals:*

1. *Passover*

2. *Shavuot*

3. *Sukkot/Shemini Arzeret/Simchat Torah*

*You are NOT interested in the weekly Sabbath (Shabbos/Shabbat) or in the High Holy Days (Rosh Hashanah/Yom Kippur).*

To enable this task, they were presented with a search engine results page (SERP) that included tags, snippets, and links to search results documents that were constructed using manual transcripts in this case.

**Task 2 – Search by keyword using SERP B**: In this task, this time users were presented with a different SERP that was

---

[1] http://www.techsmith.com/morae.html

created using automatic transcripts. In both cases, information regarding data sources (whether transcripts are manual or auto generated) was not revealed. In this case users had to search for information on the topic "*Life in the concentration camp*" amongst the displayed results. For Task 2, users were shown the SERP from "Bad" or machine-transcribed search results and given the following scenario for their information need:

*You are looking for a description regarding life in the concentration camps that may also report arrival, selection, work, the cold weather, the famine, etc.*

*Objectives for Task 1 & 2:*
- Perform A/B testing to evaluate and compare efficiency for two SERPs.
- Specifically, to analyze the results based on various criteria such as accuracy, time taken to complete the tasks, and number of clicks required before reaching the final answer.

**Task 3 – Exploring Search Engine Design**: Users were asked to go through the search interface from the homepage and to provide feedback via think-aloud observation. First, they were brought to the landing page and told to describe their overall impressions of the presentation of information architecture and state what actions they understood could be performed from the landing page. They were also asked to describe their observations of various search elements such as Suggested Searches, Popular Searches, Featured Interviews, filters based on faceted metadata, descriptions about the collection, etc. Thereafter, they were asked to perform a simple search and later asked to narrow it down based on Time Period and Location. Also, user feedback on SERP was also noted.

*Objectives:*
- Understand whether the search actions were obvious to the users
- Understand whether the users correctly interpreted the purpose of "Popular Searches" and "Suggested Searches" and could distinguish between them
- Analyze the ease with which users can complete a simple search task and check discoverability and use of filters
- Gauge effectiveness of tags and snippets while browsing through the results
- Understand if the users want to use to the search bar or the side filters to perform a search task

**Task 4 – Post-task Questionnaire**: After performing the above tasks, users completed a questionnaire rating their experience and ease of use overall and individual tasks. Feedback regarding design elements and suggestions to improve the design were captured in this digital questionnaire.

*Objectives*:
- To record user feedback on ease of performing tasks, rating helpfulness of design elements like sentence markers, speaker change markers, disfluency markers, hover maps, and tags.
- To gain insight into user reasoning while choosing either tasks as easier to perform by use of open ended questions.
- Capture suggestions to improve interface design to enable faster search of relevant results.

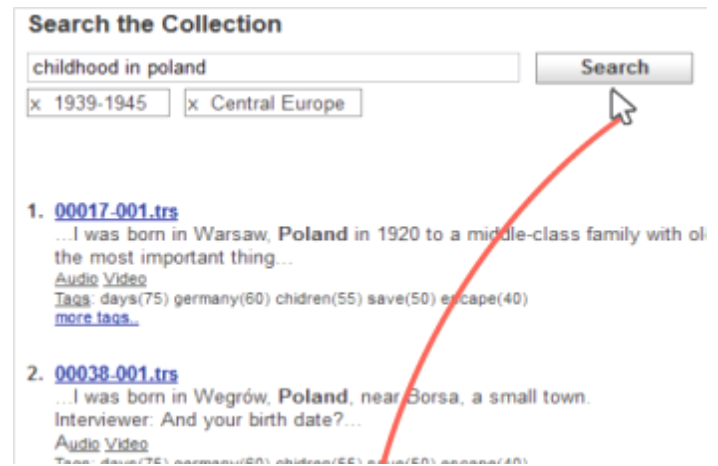

**Figure 8: Using *Morae* to capture screen and user-clicks**

## 7. RESULTS

*User Population:*

Several representative user groups with different levels of domain expertise were identified and a total of 8 users participated in this study. All these users were students living in US. Out of them, 5 were pursuing graduate degrees, and 3 pursuing bachelors. 4 users had completed high school in India, 3 in the U.S., and 1 in China. Only 3 users had prior experience researching on Holocaust but did not highly rate their familiarity with the topic. The rest of the users reported little familiarity.

*A/B Testing*:

In this study, we test two variations of the interface to perform similar task (i.e. finding relevant result). However, the users were given different query each time so that they do not use the information gained from task 1 in task 2.

Information regarding time taken to complete the task, number of clicks, and other behavior were recorded for Task 1 and Task 2. Our original design included brief summary for each search results and also the snippet in Task 2. However, while performing evaluation with the first user, we observed that user was making judgment based on summary itself while making relevance judgment and avoided going through the interview transcript. So we modified the design to remove summary from search results and modified the snippet to include sentences that appear nearby the search keywords. Hence, this evaluation includes data from the remaining users to measure accuracy and analyzing ease of performing the tasks. But, we do incorporate feedback obtained from that user during Task 3, and 4.

All 7 users were able to complete Task 1 correctly. However, for task 2, only 3 users were able to do it identify the relevant result. Figure 9 presents the percentage of accuracy achieved by comparing correct and incorrect responses obtained from users during tasks 1 and 2.

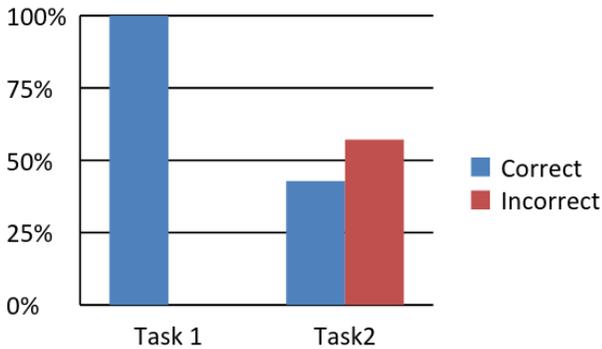

**Figure 9: Percentage accuracy in the responses**

Figure 10 presents a distribution comparing time taken by the users to complete tasks 1 and 2.

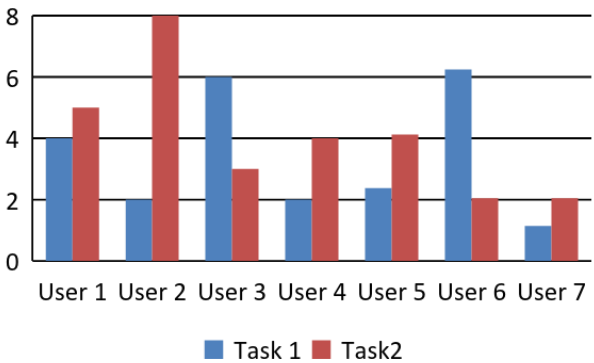

**Figure 10: Time taken to complete the task**

It can be observed that task 1 was relatively easier to complete considering 100% accuracy and relatively lesser time taken to find relevant result with the exception of 2 users. Although this perception may have been influenced by Task 1 providing key words (Passover, Sukkot, Shavuot, etc.) while Task 2 did not provide users with explicit keywords, users found the transcripts from Task 1 to be more readable and completed Task 1 with more confidence that they had selected a relevant result.

Feedback obtained from the post-task questionnaire showed that four users reported that they were able to make decisions using the snippet itself for Task 1. Whereas, for task 2 only three users reported so. All the users felt that readability of the transcript affected the time taken to find relevant search results. Two users rated Task 1 as "Quite difficult", while the rest found it of average to easy level. In case of Task 2, two users rated it as "Quite difficult", three rated it as "Average difficulty" and two found it be easy. Following features were considered helpful in determining results:

- Keyword highlighting
- Presence of tags with keyword frequency
- Readability of the transcript

According to the post-task questionnaire feedback, participants looked to keywords and tags to provide them with contextual clues about the relevancy of results. In Task 1, seeing one of the given subjects as a highlighted keyword led users to decided on the correct result. However, in Task 2, users had to infer what keywords might be from the described information need. This may led them to focus on less relevant highlighted keywords.

*Markers in Task 2 (Figure 11)*

Since transcripts used for task 2 were generated using ASR (Automatic Speech Recognition), several markers such as end of sentence markers, speaker change markers, and disfluency markers were inserted in the transcript body to enhance readability.

**End of sentence change markers**: End of sentences were marked using <s> tag in blue. Two users found these markers to be very useful, three of average usefulness, while two users did not find them helpful.

**Speaker change markers**: To distinguish between speakers (interviewer and interviewee) <spk#> tags were used in green, where # was replaced by 1 in case of interviewer and by 2 in case of interviewee. Two users found these markers to be useful, three of average usefulness, while two users did not find them helpful.

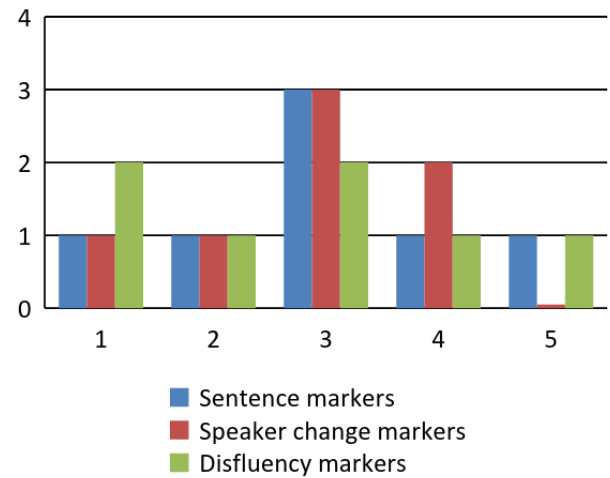

**Figure 11: Gauging helpfulness on a scale of 1-5, with 1 being "Very Unhelpful" and 5 being "Very Helpful"**

**Disfluency markers**: To highlight misspelled, repeated and other disfluencies in the text, <<text>> tags were used in dark red, for e.g. <<uhhuh>>, <<of of>>. Two users found these markers to be useful, two of average usefulness, while three users did not find them helpful.

*Exploration*:

During Exploration, participants were given a scenario stating that they were writing a research paper on the subject of "child survivors of the Holocaust." Participants were then introduced to the landing page of the prototype and asked how they would begin to find information that satisfied the specified information need. Users explored various design elements across the search engine. The purpose of "About the Collection" was deemed useful by many users, with one user mentioning it as a "great starting point for users who are new and want to know more about the collection". On asking users how they would begin to find information that satisfied their given need, their approaches diverged concerning how to begin the search process. Some users started by entering a keyword query into the search bar, while others wanted to start by selecting one of the Suggested or Popular Searches proffered on the right-hand side of the screen. Users did not tend to start with Browse By categories, although they did note these categories.

In the task to narrow down results for World War II, while most of the users immediately went to narrow their search by Time Period in the Browse By filters, some users stated that they would want to reformulate their queries in the search bar to include the keywords "World War II" and "Holocaust." When prompted for a reason, one user cited uncertainty about if it would keep the results or start an entirely new search. After expanding the Time Period filter, most users were uncertain as to which of the three year categories would yield results only for World War II, needing more information than just the numerical years to act. Users who were more familiar with World War II felt comfortable selecting the year category, 1939 - 1945, although this may have been influenced by the change-in-cursor-state when mousing over this selection, as it was the only hyperlinked text. Some also suggested providing filters such as "Pre-war", "WW II", and "Post-War" instead of year ranges. On checking the SERP with these filters, some users noticed that the number of search results had changed and confirmed that seeing fewer results was what they expected. Users also noticed the addition of the Time-Period tag filter under the search bar.

Users were then instructed to narrow their search results again, this time to include results from Poland and Germany. Users immediately looked to Location to narrow their results, including those who stated previously that they would reformulate their query. However, when they saw the options under the Location category (Eastern Europe, Central Europe, Western Europe, Scandinavia), most users were unsure of which filter would include Germany and Poland. Upon discovering the hover map, some users felt that they had enough information to choose Central Europe, while others were still uncertain and needed more explicit information. On arriving on final SERP, all users understood that they could remove one of the filters but keep the other by clicking the "x" on the tag under the search bar. When asked to comment on the presentation of search results, some users noted that the titles of the results were unhelpful and would be significantly more useful if they had a descriptive name, rather than a file name. Users also noticed the tags under each result. Some found this very helpful, while others stated that they would have to narrow their results more because the tags were very similar.

In the post-task questionnaire, users suggested several ways to improve search interface design so that the users can find relevant results in a shorter amount of time:

- Adding functionality to pin favorite results that stays on SERP while adding/removing filters
- Emphasize purpose of markers
- Including specific tags
- Having titles describing the transcripts
- Country description in the filters

## 8. FUTURE WORK

In the future, a more comprehensive representative sample would assist in accurate evaluation of the SUI designs, supported search methods, and presentation of transcription information. Future samples would include domain experts (familiar both with the subject and with using advanced search systems) such as historians and professors; subject-matter experts such as survivors, witnesses, relatives of the formers, and teachers (who may have less experience using search systems); and with domain novices with or without previous experience conducting research using advanced search systems.

Tests with such users would seek to evaluate newly incorporated elements in the SUI (resulting from the current evaluative study) such as more specific labeling and categories and more intuitive interactions such as for facet selection and results refinement. Additionally, researchers would utilize more traditional A/B testing to evaluate different search elements. In the current study, performing separate study for task 1 and 2 may have better addressed the question of efficiency. Though it would have required more number of users and randomizing interface presentation, and since this study involved only 7 users, it was not possible to complete it in present scope of study.

Additionally, future studies with Morae will benefit from a more comprehensive set of metrics by which to evaluate efficiency, the presentation of information, and search behaviors of users. Utilize Morae's native tools will provide excellent means of comparison between user results.

Possible hypotheses for the future might center around testing different in-text automatic transcription markers, whether placing more visual emphasis on the providing a key to speech markers with the transcripts would affect how users perceived the markers, and if changing the tag presentation in search results from just frequency to something more specific may improve users' selection of relevant results.

## 9. CONCLUSION

We were able to verify and test our research hypotheses in this study. It was hypothesized in H1 that users would prefer human-transcribed search results, as opposed to automatically generated ones. From our A/B testing, we obtained 100% accuracy in Task 1 vs. 42.8% in Task 2 that helped us conclude that H1 holds true and readability is an important factor aiding users in decision making. Obtaining feedback from users on different markers (end-of-sentence, disfluency, speaker changes) we observed that disfluency markers were quite helpful with users reporting that they would prefer if these highlighted disfluencies can be removed altogether to enhance readability and would not lead to loss of information. But feedback on end-of-sentence markers showed ambiguous results; hence H2 remains inconclusive with the present evaluation. Against our expectations and hypothesis H3, users reported that marking speaker changes did not improve understanding of the underlying text nor did it increase efficiency to perform the search tasks. We were also able to have clear understanding on users' perspectives on the design elements and ease of parsing information. Overall user experience can be improved by adequate use of UI elements, like highlighted search items, tags, snippets, etc., to promote discovery and exploratory search. These interface elements well supported the search task, and helped in query reformulation and narrowing search results. For novice users, presenting popular and suggested search along with brief description about the data collection can serve as a starting point for future searches.

# 11. APPENDIX

This section includes screenshots of our search engine showing different design elements.

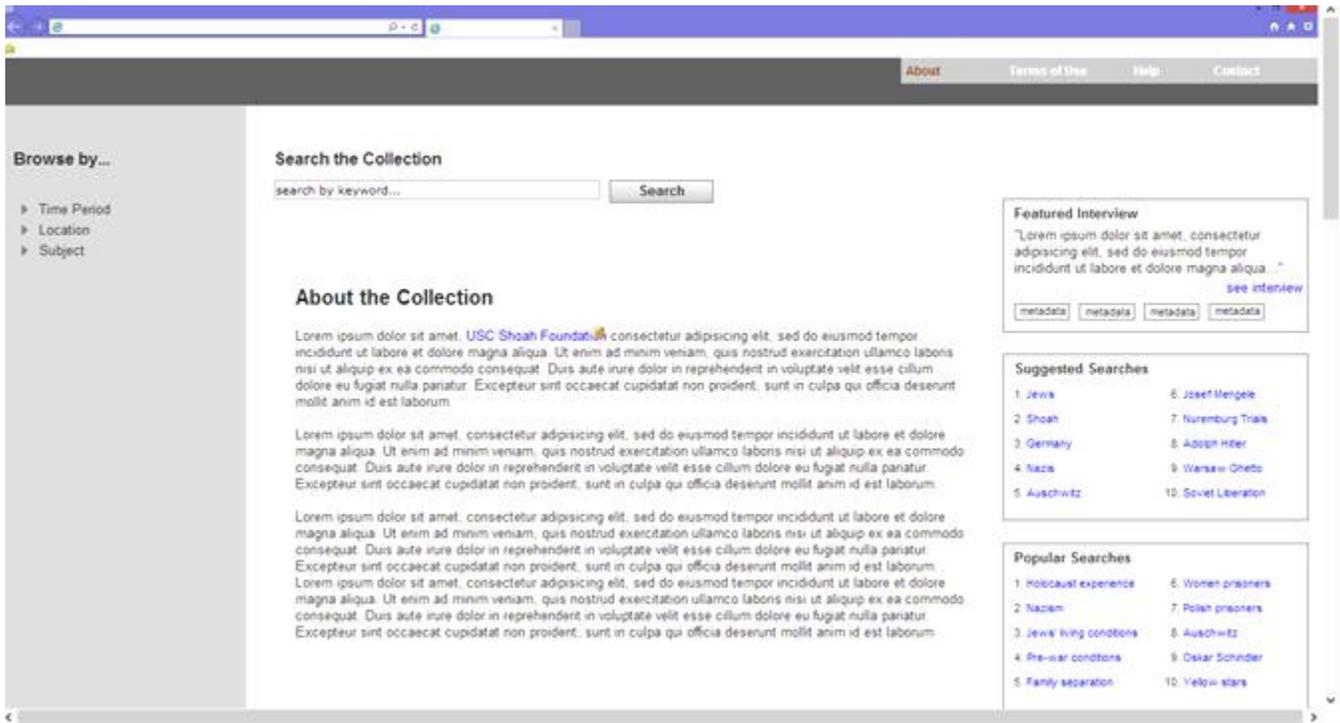

**Figure 12: Landing Page**

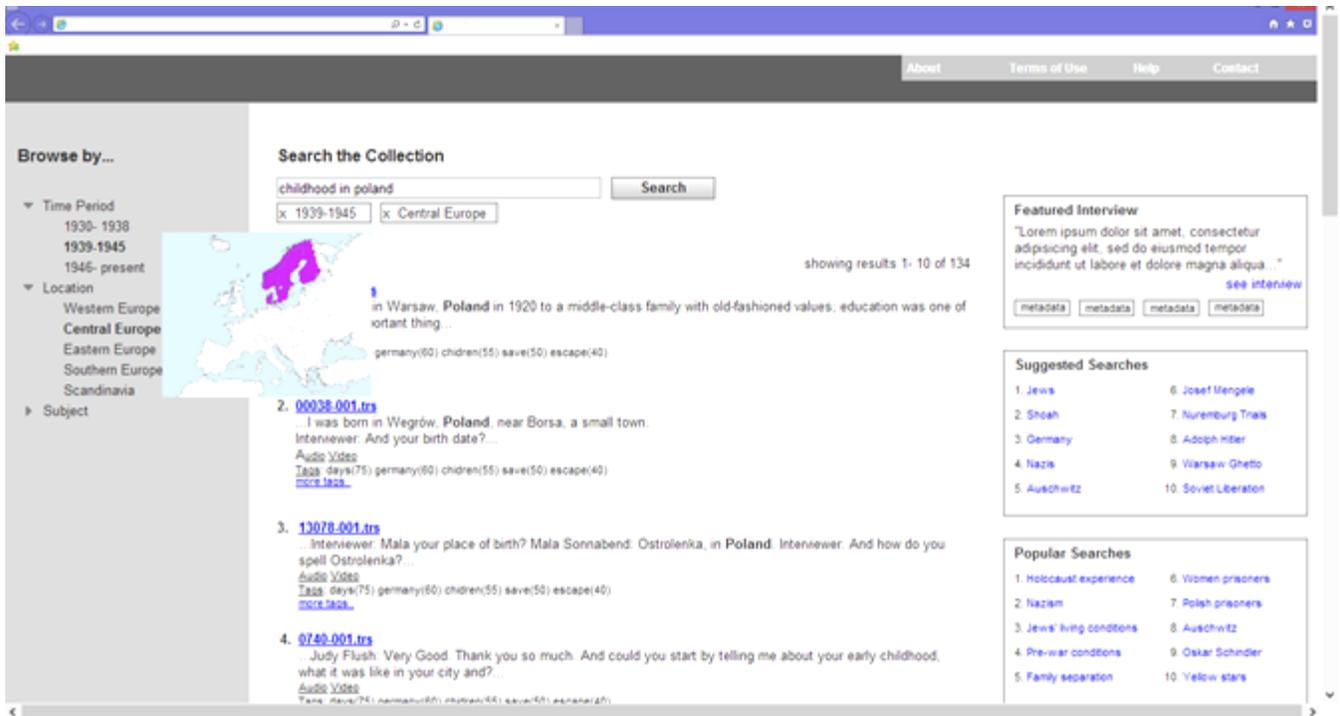

**Figure 13: Search Results Page with filters on the right, search box & filter tags in top-center, and feature interview, suggested searches, popular searches on the right**

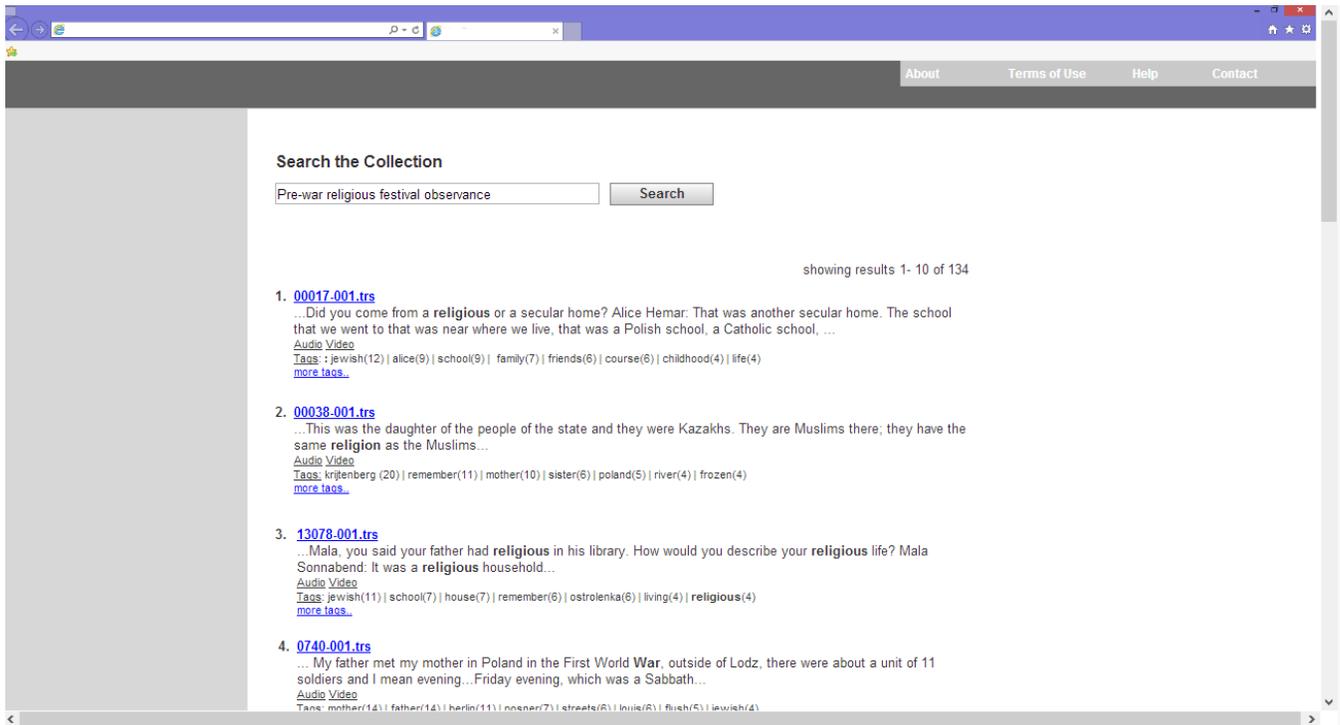

Figure 14: SERP A with manual transcripts

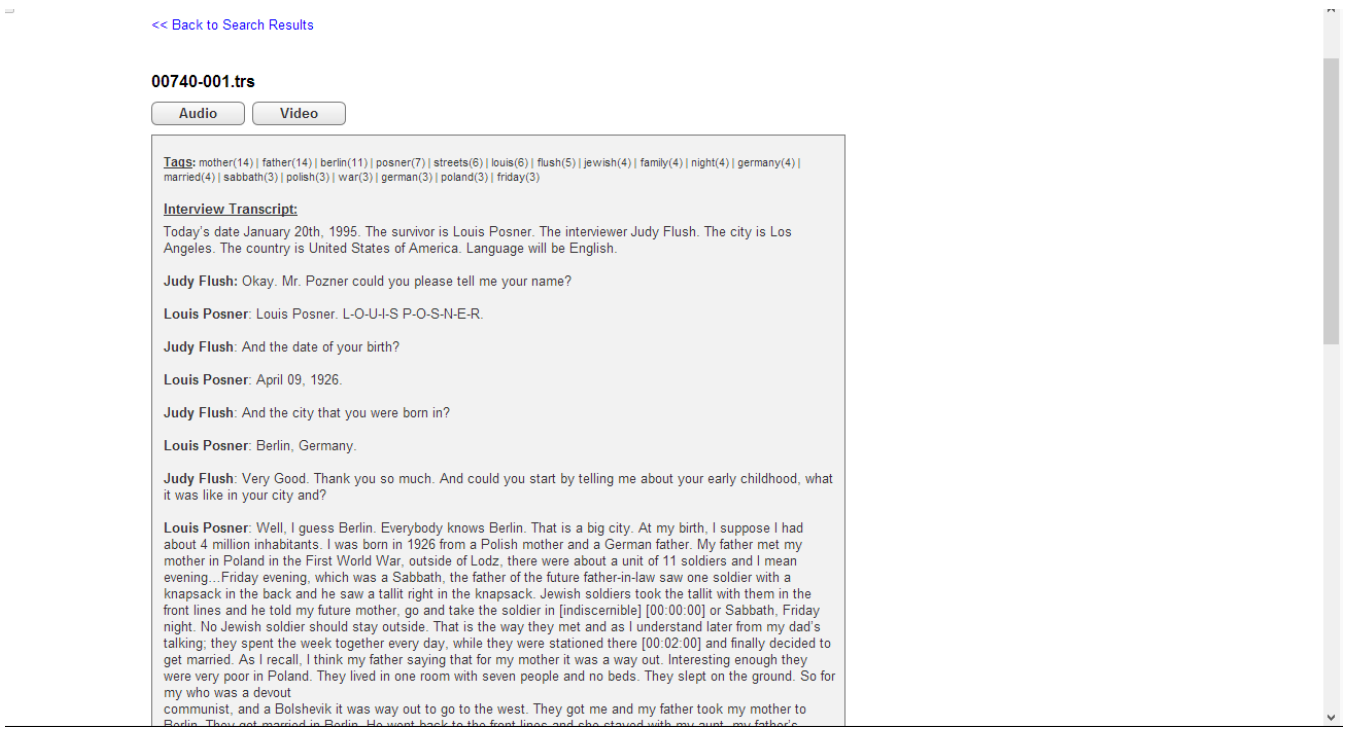

Figure 15: Sample Search result from SERP A

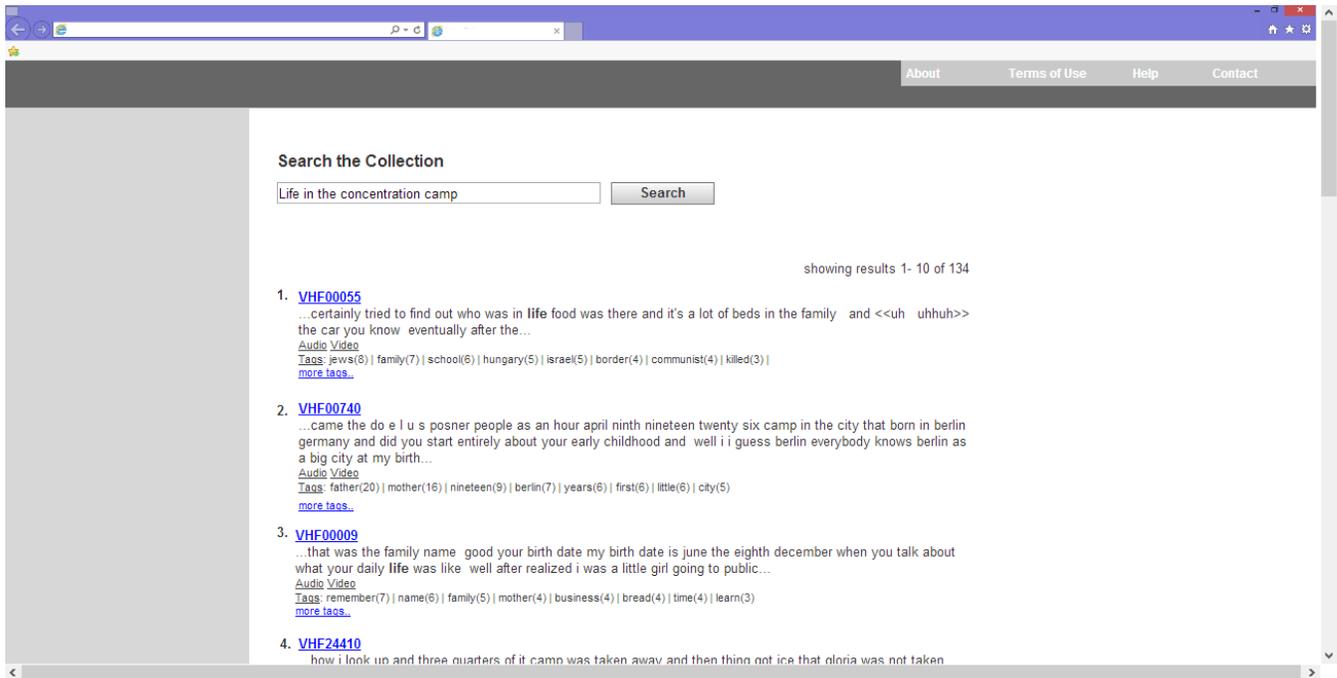

Figure 16: SERP B with automated transcripts

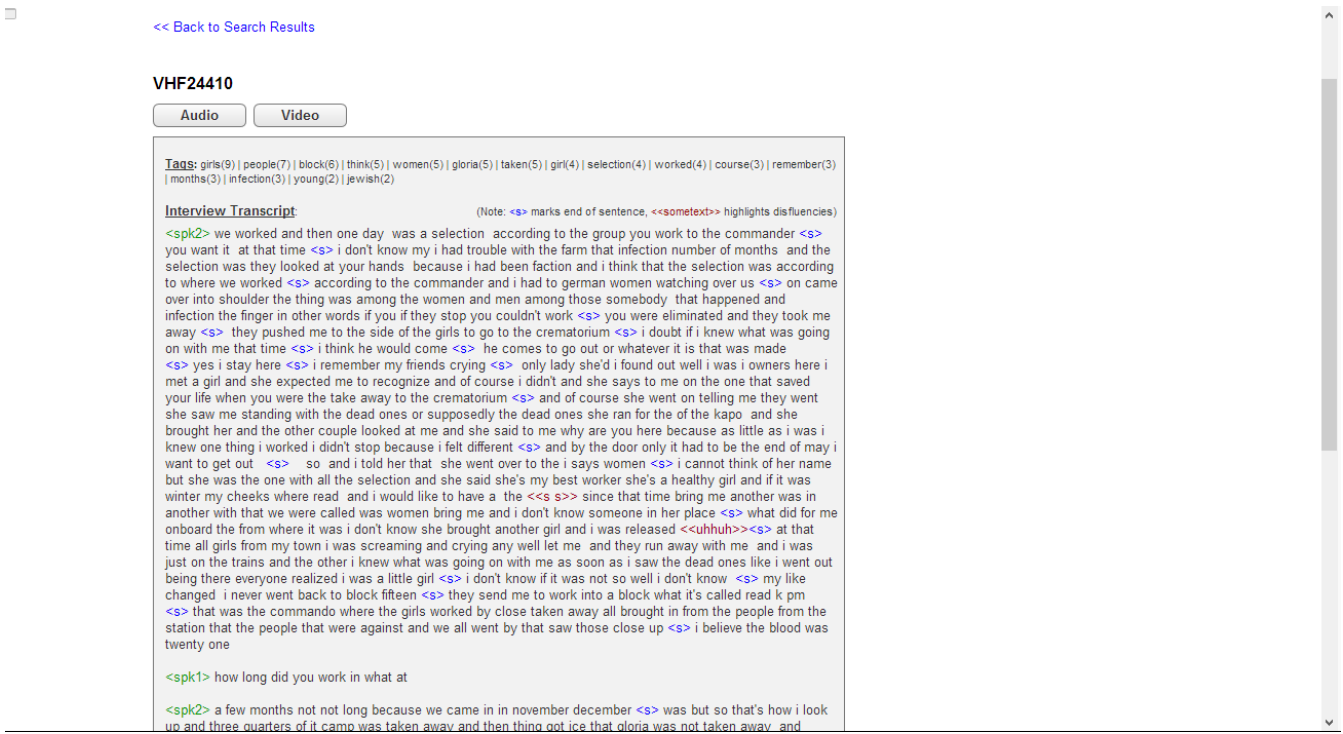

Figure 17: Sample Search Result from SERP B